\documentclass[a4paper,11pt]{article}
\usepackage{amssymb,rotating,graphics,epsfig,float,graphics}
\usepackage{amsmath,verbatim,a4wide}
\usepackage{theorem}

\theoremstyle{break} 
{\theorembodyfont{\rmfamily} \newtheorem{ex}{Example}[section]}
{\theorembodyfont{\rmfamily} }

{
\begin{document}

\author{Ferdinand Verhulst \\
University of Utrecht, Department of mathematics \\
PO Box 80.010, 3508 TA Utrecht, The Netherlands 
}
\title{ Variations on the Fermi-Pasta-Ulam chain, a survey}

%\date{  }
\maketitle

MSC classes:	37J20, 37J40, 34C20, 58K70, 37G05, 70H33, 70K30, 70K45\\

Key words: Fermi-Pasta-Ulam, resonance, periodic solutions, normalisation, chaos, symmetry, 
Hamilton-Hopf bifurcation. 

\begin{abstract} 
We will present a survey of low energy periodic Fermi-Pasta-Ulam chains with leading idea the  
"breaking of symmetry". The classical periodic FPU-chain (equal masses for all 
particles) was analysed by Rink in 2001 with main conclusions that the normal form 
of the beta-chain is always integrable and that in many cases this also holds for 
the alfa-chain. The implication is that the KAM-theorem applies to the classical chain 
so that at low energy most orbits are located on invariant tori and display quasi-periodic 
 behavior. Most of the reasoning also applies to the FPU-chain with fixed endpoints.

The FPU-chain with alternating masses already shows a certain breaking of symmetry. 
Three exact families of periodic solutions can be identified and a few exact invariant manifolds which are  
related to the results of Chechin et al.~(1998-2005) on bushes of periodic solutions. 
An alternating 
chain of 2n particles is present as submanifold in chains with k 2n particles, k=2, 3, ...
The normal forms are strongly dependent on the alternating masses 1, m, 1, m,... If m is not equal to 
2 or 4/3 the cubic normal form of the Hamiltonian vanishes.  For alfa-chains there are some 
open questions regarding the integrability of the normal forms if m= 2 or 4/3. Interaction between the optical 
and acoustical group in the case of large mass m is demonstrated.  
 
The part played by resonance suggests the role of the mass ratios. It turns out that 
in the case of 4 particles there are 3 first order resonances and 10 second order ones; 
the 1:1:1:...:1 resonance does not arise for any number of particles and mass ratios. An interesting case is 
the 1:2:3 resonance that produces after a Hamilton-Hopf bifurcation and breaking symmetry chaotic behaviour in 
the sense of Shilnikov-Devaney. Another interesting case is the 1:2:4 resonance. 
As expected the analysis of various cases has a significant impact on recurrence phenomena; 
this will be illustrated by numerical results.  
\end{abstract}

 \section{Introduction } \label{chain} 
\begin{figure}[ht]
 \begin{center}
  \resizebox{6cm}{!}{
   \includegraphics{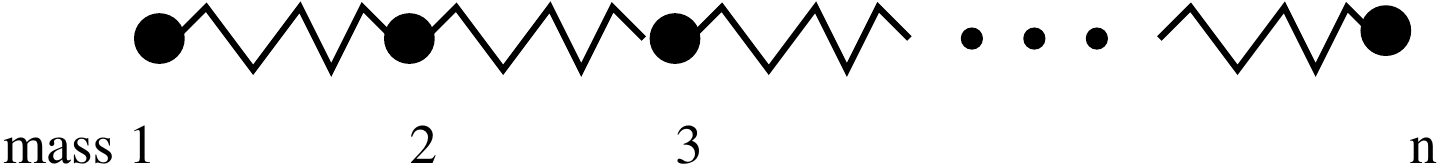}} 
  \end{center}
 \caption{A Fermi-Pasta-Ulam chain with fixed endpoints.} 
\label{FPUpict}
\end{figure}

 Chains of oscillators arise naturally in systems of coupled oscillators and by discretisation of vibration problems 
 of structures. In physics studying the Fermi-Pasta-Ulam (FPU) chain has been very influential for a different reason. The 
 FPU-chain models a one-dimensional chain of oscillators with nearest-neighbour interaction only; 
 see fig.~\ref{FPUpict}. It was formulated 
 to show the thermalisation of interacting particles by starting with exciting one mode with the expectation that after 
 some time the energy would spread out over all the modes. This is one of the basic ideas of statistical mechanics.
  In the first numerical experiment in 1955,  32 oscillators  were used with 
 the spectacular outcome that the dynamics was recurrent as after some time most of the energy returned to the 
 chosen initial state. For the original report see Fermi et al.~\cite{FPU55} and a review by Ford~\cite{Ford}, 
 recent references can be found in Christodoulidi et al.~\cite{CEB} or Bountis and Skokos \cite{BS12}.  
 Discussions can be found in Jackson~\cite{J91}, Campbell et al.~\cite{CRZ05} and Galavotti (ed.)\cite{G08}. 
 Note that although studies of FPU-chains are of great interest, as models for statistical mechanics 
 problems they are too restrictive. 

\subsection{Formulation}
 
 The original FPU-chain was designed with fixed endpoints and choosing the initial energy small. Later research showed the presence 
 of periodic solutions and wave phenomena, also larger values of the energy were considered. Another version of the 
 FPU-chain is the spatially 
 periodic chain where particle 1 is connected with the last one. In this survey we will focus mainly on the periodic chain 
 with small initial values of the energy. The Hamiltonian $H(p, q)$ for $N$ particles is of the form: 
 \begin{equation} \label{Hfpu}
 H(p, q) = \sum_{j=1}^N \left( \frac{1}{2m_j}p_j^2 + V(q_{j+1} - q_j) \right), 
 \end{equation} 
 where particle 1 is connected with particle $N$. The coordinate system has been chosen so that $q= p=0$ is a 
 stable equilibrium. For FPU-chains one considers usually potentials $V(z)$ that contain quadratic, cubic and quartic terms.
 Explicity 
 \[ V(z) = \frac{1}{2}z^2 + \frac{\alpha}{3} z^3 + \frac{\beta}{4}z^4. \] 
 If $\beta =0$ we call the FPU-chain an $\alpha$-chain, if $\alpha =0$ a $\beta$-chain. Physically the 2 chains are different, 
 for an $\alpha$-chain the forces on each particle are asymmetric, for  a $\beta$-chain they are symmetric.\\ 
 The spatially periodic chain has a second integral of motion, the momentum integral:
 \begin{equation} \label{mom}
 m_1 \dot{q}_1 + m_2 \dot{q}_2  + \ldots + m_N \dot{q}_N  = \,{\rm constant}. 
 \end{equation} 
 The momentum integral \eqref{mom} enables us to reduce the $N$ dof system to a $N-1$ dof Hamiltonian 
 system by a symplectic transformation. \\
For low energy orbits near stable equilibrium one usually rescales $p \mapsto \varepsilon \bar{p}, q \mapsto \varepsilon \bar{q}$, divides the 
 Hamiltonian by $\varepsilon^2$ and drops the bars. For the linearised system near stable equilibrium we find: 
 \begin{eqnarray} \label{FPUlin} 
\begin{cases} 
m_1 \ddot{q}_1 + 2q_1 -q_2 - q_N & =  0,\\ 
m_2 \ddot{q}_2 + 2q_2 -q_3 - q_1 & =  0,\\ 
m_3 \ddot{q}_3 + 2q_3 -q_4 - q_2 & =  0,\\ 
\ldots  & =  0,\\
m_N \ddot{q}_N + 2q_N -q_1 - q_{N-1} & =  0.
\end{cases}
\end{eqnarray} 
The quadratic nonlinearities start with $\varepsilon$, the cubic ones with $\varepsilon^2$. 
 The spectrum of the linear operator (the eigenvalues near stable equilibrium) determines the resonances and the
 nonlinear dynamics  near stable equilibrium. Our survey is based on papers that make extensive use of 
 normalisation-averaging techniques, 
 see Sanders et al.~\cite{SVM}, chs.~2 and 10. This involves near-identity transformations to simplify the equations of motion 
 or the Hamiltonian itself if one studies such a system. A quadratic Hamiltonian indicated by $H_2$ corresponds 
 with a linear system of differential equations; 
 for a Hamiltonian with cubic terms near-identity transformation removes the non-resonant terms to higher order. 
 Omitting the higher order terms the resulting normalised Hamiltonian $\bar{H}=H_2 + \bar{H}_3$ contains only 
 the resonant terms $\bar{H}_3$ of the cubic $H_3$ (the index indicates the power of the polynomials). 
 One can go on with the normalisation proces by using a near-identity transformation to remove the non-resonant terms 
 from $H_4$, etc. 
 
 In general the normalised (averaged) equations that are truncated at some level of normalisation will not be integrable, 
 although there are many exceptions. For the FPU-Hamiltonian in homogeneous polynomials we have the notation:
 \[ H = H_2 + \varepsilon H_3 + \varepsilon^2 H_4,\,\, {\rm and}\,\, \bar{H} = H_2 + \varepsilon \bar{H}_3 + 
 \varepsilon^2 \bar{H}_4. \]

 We will describe a number of prominent cases that show different dynamics for different choices of the masses. 
 In the original (classical) FPU problem all masses are equal which seems a natural choice. A second natural choice is to alternate 
 the masses $m, M, m, M, \ldots, m, M$; it is no restriction to assume $0< m \leq M$. A quite different approach 
 is to look for mass ratio's that produce interesting 
 resonances and dynamics. We aim at summarising all these approaches for low energy chains. 
  Of special interest in the analysis are integrals corresponding with approximate invariant manifolds of the 
 averaged systems, periodic solutions, bifurcations and chaos. \\
 {\em An important conclusion will be that the classical FPU-chain contains so  many symmetries that
 by symmetry breaking it is structurally unstable}.
 
 \subsection{Theoretical background}
 There exist an enormous amount of papers on the original FPU-chain of Fermi et al.~\cite{FPU55}. 
 A large number of the papers 
 consist of numerical explorations; they are often inspiring but not always satisfactorily explaining the phenomena. 
 Apart from normalisation-averaging, symmetry considerations are important for the  qualitative results. 
 This involves the theory of Hamiltonian systems, see for an introduction Verhulst \cite{FV85}  and for the more 
 general dynamical 
 systems context Broer et al.~\cite{BT}. New results on Hamiltonian systems and symmetry are found in  
 Bountis and Skokos \cite{BS12},  Efstathiou \cite{E05} and  Han{\ss}mann \cite{H07}.  
 Basic understanding of recurrence as formulated by 
 Poincar\'e~\cite{PMC} vol.~3, ch.~26 is essential.\\  
 A systematic study of dynamical systems with discrete symmetry was started by Chechin and Sakhnenko \cite{CS98}. 
 The authors introduce 
 the notion of {\em bushes}  with a bush comprising all modes singling out an active symmetry group in the system. A 
 bush corresponds with a lower dimensional invariant manifold (or approximate invariant manifold in the sense of 
 normalisation) giving insight in the various dynamical parts that compose the system. The theory is quite general, 
 it was applied to FPU chains by Chechin et al.~in \cite{CNA} and \cite{CRZ}. \\ 
 Independently the ideas of utilising symmetries were also developed by Rink~\cite{BR} and by 
 Bruggeman and Verhulst in \cite{BV17} and \cite{BValt}.

 \section{The classical periodic FPU-chain} \label{classfpu}
 In the original FPU problem one considered the so-called mono-atomic case, i.e. all masses equal; we call this the 
 classical FPU-chain and put $m_1=m_2= \ldots = m_N=1$. The recurrence of the classical FPU-chain signalled by Fermi et 
 al.~\cite{FPU55} was surprising at the time as this was before the time of publication of the KAM theorem (see below).\\ The 
 linearised system \eqref{FPUlin} has the frequencies $\omega_j$ of the corresponding harmonic equations: 
 \begin{equation} \label{eigclas} 
 \omega_j = 2 \sin \left(\frac{j \pi}{N} \right), \,j=1, \ldots, N.
 \end{equation} 
 The implication is that we have many $1:1$ resonances,  $N/2$ if $N$ is even and $(N-1)/2$ if $N$ is odd. Also there exist 
 accidental other resonances like $1:2:1$. A natural first step is to reduce the system using integral \eqref{mom} to $N-1$ 
 dof. \\ 
 An interesting attempt to solve the recurrence problem was made by Nishida \cite{N71} 
 by proposing to use the KAM theorem; this theorem guarantees under the right conditions the existence of an infinite 
 number of $(N-1)$-tori containing  quasi-periodic solutions near stable equilibrium. This would solve the recurrence 
 problem, but unfortunately the spectrum is resonant and the KAM theorem can not be applied in a simple way.\\ 
 The problem was for most cases solved for the spatially periodic FPU chain by Rink in \cite{BR}; 
 his results can also be applied to the chain 
 with endpoints fixed. We summarise the reasoning. First the system with cubic and quartic terms in the Hamiltonian is 
 transformed by symplectic normalisation (also called Birkhoff-Gustavson normalisation) to a simpler form. If the 
 resulting normalised Hamiltonian $\bar{H}$ is nondegenerate in the sense of the KAM theorem and if it is integrable i.e 
 containing, in addition to integral \eqref{mom},  $N-1$ functionally independent integrals that are in involution, 
 then the KAM theorem applies to the original 
 Hamiltonian $H$. By the transformation the nonresonant terms of the cubic and quartic part are shifted to higher order. 
 The original system contains various discrete symmetry groups, a rotation symmetry and a reflection symmetry. 
 These symmetries carry over to the normalised Hamiltonian system with the surprising result that the cubic terms in 
 $\bar{H}$ vanish!  From theorem 8.2 of Rink \cite{BR} we have for the classical periodic FPU chain derived from Hamiltonian 
 \eqref{Hfpu} containing cubic and quartic terms:
 \begin{equation} \label{H3}
 \bar{H}_3 = 0.
 \end{equation} 
 The analysis in Rink \cite{BR} of $\bar{H}$ produces furthermore: 
 \begin{enumerate} 
 \item Assume $\alpha \neq 0$ and $N$ is odd, then $H_2 + \varepsilon^2 \bar{H}_4$ is integrable and 
 nondegenerate in the sense of the KAM theorem.
 \item Assume $\alpha \neq 0$ and $N$ is even, then $H_2 + \varepsilon^2 \bar{H}_4$ hast at least $(3N-4)/4$ 
 quadratic integrals (if 4 
 divides $N$) or $(3N-2)/4$ quadratic integrals (if 4 does not divides $N$). 
 \item  The normalised $\beta$-chain ($\alpha =0$) is integrable and nondegenerate in the sense of the  KAM theorem. 
 Almost all low-energy orbits are periodic or quasi-periodic and move on invariant tori near stable equilibrium. 
 \item Similar results can be obtained for the classical FPU-chain with fixed endpoints. 
 \end{enumerate} 
 The remaining problem is the integrability of $H_2 + \varepsilon^2 \bar{H}_4$ in the case of the even $\alpha$-chain. 
 To check 
 this one has to carry out the normalisation to quartic terms which is quite a lot of work if $N$ is large. We will discuss 
 an example with $\alpha =1, \beta = -1$. 
  \begin{ex} 
 Consider a periodic Fermi-Pasta-Ulam chain consisting of four particles of equal mass m (= 1) with quadratic
and cubic nearest-neighbor interaction. Periodic means that we connect the first with the fourth particle. The Hamiltonian is 
in this case: 
\begin{equation} \label{FPU4} 
H(p, q)= \sum_{j=1}^4 (\frac{1}{2}p_j^2 + V(q_{j+1}-q_j)),
\end{equation}
 with 
 \[ V(z)= \frac{1}{2}z^2 + \frac{1}{3}z^3 - \frac{1}{4}z^4.  \]
 The corresponding equations of motion were studied  Rink and Verhulst \cite{RV00}.

\begin{figure}[ht]
 \begin{center}
  \resizebox{10cm}{!}{
   \includegraphics{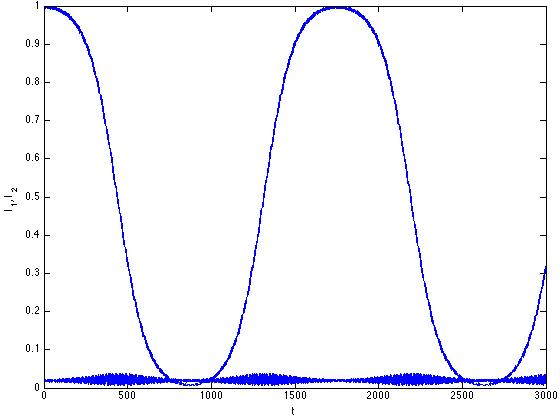} \,  \includegraphics{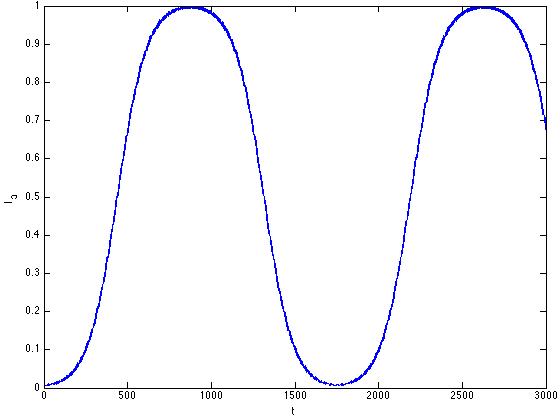}} 
  \end{center}
 \caption{The actions for 3000 timesteps near the unstable $x_2$ normal mode of system \eqref{FPU4eqs} with
$\varepsilon = 0.1$, initial conditions $x_1(0) = x_3(0) = 0.1, x_2(0) = 1$ and initial velocities zero. Left the
action $I_2(t) = \frac{1}{2} (\dot{x}_2^2 + 2x_2^2)$ starting near zero and increasing to values near 1; also the 
nonresonant $I_1(t) = \frac{1}{2} (\dot{x}_1^2 + 4x_1^2)$. 
Right the resonant action $I_3(t) = \frac{1}{2} (\dot{x}_3^2 + 2x_3^2)$) that exchanges energy with the $x_2$ mode 
(pictures from \cite{FV18}).} 
\label{FPUfig}
\end{figure} 
The equations induced by Hamiltonian \eqref{FPU4} have a second integral of motion, the momentum integral 
$ \sum_{j=1}^4p_j =$ constant. This enables us to reduce the 4 dof equations of motion to 3 dof by a canonical  (symplectic)  
transformation.  From Rink and Verhulst \cite{RV00} we have the reduced system:
\begin{eqnarray} \label{FPU4eqs} 
\begin{cases} 
\ddot{x}_1 + 4x_1 &= 4x_2x_3 + 4x_1^3 + 6x_1(x_2^2 +x_3^2),\\ 
\ddot{x}_2 + 2x_2 &= 4x_1x_3 + x_2^3 + 3x_2(x_3^2 +2x_1^2),\\ 
\ddot{x}_3 + 2x_3 &= 4x_1x_2 + x_3^3 + 3x_3(x_2^2 +2x_1^2).
\end{cases}
\end{eqnarray} 
We can identify 3 families of periodic solutions, the 3 normal modes in the coordinate planes. Consider the $x_2$ normal 
mode that satisfies the equation: 
\[ \ddot{x}_2 + 2x_2 = x_2^3. \]
 In general, solutions far from stable equilibrium become chaotic, so we restrict ourselves to a neighbourhood of the origin 
 by rescaling $x_1= \varepsilon \bar{x}_1, x_2= \varepsilon \bar{x}_2, x_3= \varepsilon \bar{x}_3$ and then omitting the bars. 
 Rescale also $\sqrt{2}t =s$. System \eqref{FPU4eqs} becomes: 
 \begin{eqnarray} \label{FPU4eqs2} 
\begin{cases} 
\frac{d^2x_1}{ds^2} + 2x_1 &= 2 \varepsilon x_2x_3 + 2 \varepsilon^2x_1^3 + 3 \varepsilon^2x_1(x_2^2 +x_3^2),\\ 
\frac{d^2x_2}{ds^2} + x_2 &= 2 \varepsilon x_1x_3 + \frac{1}{2} \varepsilon^2 x_2^3 + 
\frac{3}{2} \varepsilon^2x_2(x_3^2 +2x_1^2),\\ 
\frac{d^2x_3}{ds^2}+ x_3 &= 2 \varepsilon x_1x_2 + \frac{1}{2} \varepsilon^2 x_3^3 + 
\frac{3}{2} \varepsilon^2x_3(x_2^2 +2x_1^2).
\end{cases}
\end{eqnarray} 
The equation for the $x_2$ normal mode was studied in many introductions to the averaging method, where with 
initial  values 
$x_2(0)=a, dx_2(0)/ds=0$ we obtain the approximation:
\[ \phi(s) = a \cos(s - \varepsilon^2\frac{3}{16}a^2 s). \] 
We transform $x_1=y_1,  x_2= \phi(s) +y_2, x_3=y_3$ in system \eqref{FPU4eqs2} and linearising we  find:  
 \begin{eqnarray} \label{FPU4eqslin} 
\begin{cases} 
\frac{d^2y_1}{ds^2} + 2y_1 &= 2 \varepsilon \phi(s)y_3  + 3 \varepsilon^2y_1\phi^2(s),\\ 
\frac{d^2y_2}{ds^2} + y_2 &= 0,\\ 
\frac{d^2y_3}{ds^2}+ y_3 &= 2 \varepsilon y_1\phi(s) + 
\frac{3}{2} \varepsilon^2y_3 \phi^2(s).
\end{cases}
\end{eqnarray} 
The first and third equations are coupled but there is no resonance because of the basic frequencies $\sqrt{2}$ and 1; we 
conclude that the solutions of system \eqref{FPU4eqslin} are stable. Interestingly, it was proved in 
\cite{RV00} that near stable equilibrium the stability in linear approximation is destroyed by the nonlinearities. 
See fig.~\ref{FPUfig} for an illustration. 
 \end{ex}  
 
 \begin{ex} 
 {\bf Other examples}\\ 
 The relatively simple case of 3 particles was discussed by Ford \cite{Ford}; the system is identified with the H\'enon-Heiles system, a 2 dof Hamiltonian system in $1:1$ resonance; for a survey see Rod and Churchill \cite{RC85}.  
This is interesting as this system has an integrable normal form for low energy values. Between the invariant tori there 
 exists chaos but of exponentially small measure. If the energy is increased the amount of chaos increases, destroying more 
 and more  tori until the system looks fully chaotic at higher energy. Proofs are available for this behaviour, see Holmes et al.~\cite{H88}, 
 except that we do not know whether at `'full chaotic 
 behaviour`` there are no tiny sets of tori left, undetected by numerics.\\ 
 In Rink and Verhulst \cite{RV00} the classical system with 4, 5 and 6 particles was analysed in the cases of $\alpha$- and 
 $\beta$-chains, also
 for mixed cubic  and quartic terms. In these examples the normal forms are integrable. 
 \end{ex} 
 
 \section{The FPU-chain with alternating masses} 
 Alternating the masses of a FPU-chain produces already a certain symmetry breaking. It is no restriction to rescale 
 the smallest mass to 1 and have largest mass $m \geq 1$.\\ 
 So we consider the periodic FPU-chain with $N$ (even) masses that alternate: $1, m, 1, m, \ldots, 1, m$  
 (the case $0 <m \leq 1$ follows from symmetry considerations). The chain is related to
 the formulation in Galgani et al.~\cite{GGMV} that analyses the chain and explores numerical aspects if $N$ is large. 
 In Bruggeman and 
Verhulst \cite{BValt} a general analysis was started, but 
 there are still many open questions; we summarise a number of results of this paper. \\ 
 The eigenvalues $\lambda_j, j= 1, \ldots N$ of system \eqref{FPUlin} are with $a=1/m$ in the case of alternating masses: 
 \begin{equation} \label{eigalt} 
 \lambda_j = 1+a \pm  \sqrt{1+ 2a \cos(2 \pi j/N) +a^2}, j = 1, \ldots N. 
 \end{equation} 
 Several observations can be made: 
 \begin{enumerate} 
 \item One eigenvalue equals 0 corresponding with the existence of the momentum integral \eqref{mom}. 
 
 \item If $N$ is  a multiple of 4 we have among the eigenvalues the numbers $2(a+1), 2, 2a$. 
 
 \item For large masses $m$ ($a \rightarrow 0$) the eigenvalue spectrum consists of 2 groups, one with size $2+O(a)$ 
 (the so-called optical group) and one with size $O(a)$ (the so-called acoustical group). The symplectic transformation to 
 $N-1$ dof mixes the modes because of the nearest-neighbour 
 interactions, present already in the linearised system \eqref{FPUlin}. So we cannot simply identify the dynamics of the optical group 
 with the dynamics of the large masses. 
 
 \end{enumerate} 
 A few qualitative and quantitative results were obtained by  Bruggeman and Verhulst \cite{BValt}: 
 \begin{enumerate} 
 \item 
 We can identify {\em three explicit families of periodic solutions} characterised by the frequencies $\sqrt{2}, \sqrt{2a}, 
 \sqrt{2(1+a)}$. The solutions are either harmonic or elliptic functions. 
 \item 
 In the spirit of Chechin and Sakhenko \cite{CS98} we can identify {\em bushes} of solutions in the following sense: 
 the dynamics of a system with 
 $N$ particles will be found as a submanifold in systems with $kN$ particles ($k =2, 3, \ldots$). This increases the importance 
 of studying chains with a small number of particles enormously. Note that the result is valid for large values of $N$, 
 it also holds in the classical case $m=1$.
 \item 
 First order averaging-normalisation ($m \neq 1$) produces for the $\alpha$-chain only non-trivial results if $m=2$ and 
 $m= 4/3$. From the point of view of normalisation 
 the case of large $m$ ($a \rightarrow 0$) has to be treated separately. 
 \item An interesting discussion by Zaslavsky \cite{Z} deals with the phenomenon of delay of recurrence in Hamiltonian systems by 
 {\em quasi-trapping}. This phenomenon arises for 3 and more dof if resonance manifolds, acting as subsets of the 
 energy manifold, contain periodic solutions surrounded by invariant tori. The orbits entering such resonance manifolds may 
 be delayed passage by staying for a number of revolutions near these tori.\\
 In Bruggeman and Verhulst \cite{BValt} an explicit analysis and numerics of quasi-trapping is given for a number of 
 cases with 8 particles. In the case of large mass $m$ a second order normalisation is necessary; the recurrence is 
 sensitive to the initial conditions.
 
 \item For the alternating mass $m$ large (small $a$) we expect different dynamics for the optical group (eigenvalues near 2) and 
 the acoustical group (eigenvalues $O(a)$), see Galgani et al.~\cite{GGMV}.  
 This raises an old question: can high frequency modes transfer energy to low frequency modes and vice versa? The answer is affirmative, 
 see the discussion below and fig.~\ref{FPUfig2}.    
 \end{enumerate} 
 
We summarise results for the cases $N=4n$ and $N=8n$.

\subsection{Chain with 4n particles, $n = 1, 2, 3, \ldots$, \cite{BV18}}
A system with 4 particles is imbedded as an invariant manifold in a system with 4n particles. 
The momentum integral \eqref{mom} enables reduction to 3 dof with frequencies $\sqrt{2}, \sqrt{2a},  \sqrt{2(1+a)}$. 
We find no 3 dof first order resonances in a system with 4 particles. 
The normal modes are exact periodic solutions both for the $\alpha$- and the $\beta$-chain. The normal forms are in 
both cases integrable to second order. The recurrence of the orbits on an energy manifold depends on the initial conditions, 
starting near an unstable periodic orbit lengthens the recurrence times. \\
For the case large mass $m$ ($a$ small) see below.

\subsection{Chain with  8n particles, $n = 1, 2, 3, \ldots$, \cite{BValt}} 
A system with 8 particles is imbedded as an invariant manifold in a system with 8n particles. 
Using integral \eqref{mom} produces reduction to 7 dof with frequencies: 
\[ \sqrt{2}, \sqrt{2a},  \sqrt{2(1+a)}, 1+ a + \sqrt{1+a^2}\, ({\rm twice}), 1+ a - \sqrt{1+a^2}\, ({\rm twice}). \] 
The normal forms become much more complex ($H_4$ contains 49 terms) so we restrict the analysis to $\alpha$-chains. 
As expected we recover the invariant manifold associated with the first 3 eigenvalues (or frequencies) for the system 
{\em before} normalisation; we find two more 6-dimensional invariant manifolds of the exact equations. The 3 invariant 
manifolds have the normal mode  periodic solution associated with the frequency $2a$ in common. This mode plays 
a pivotal part in the dynamics.   \\
Normalisation produces $\bar{H}_3 =0$ except if $a= 0.5, 0.75$ and if $a$ is close to zero (large mass). The normal form flow 
in the 3 invariant manifolds is integrable. In the case $a= 0.75$ we find instability of the invariant manifolds, the stability in 
the other cases can not be decided as the eigenvalues are purely imaginary (this is a basic stability problem of Hamiltonian systems 
with more than 2 dof). \\
A conclusion is that the presence of nested invariant manifolds (bushes) makes the equipartition of energy rather improbable. 

\subsection{Interactions between optical and acoustical group} 
\begin{figure}[ht]
 \begin{center}
  \resizebox{14cm}{!}{
   \includegraphics{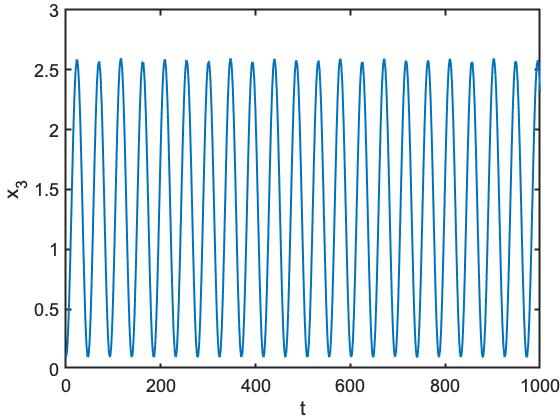} \,  \includegraphics{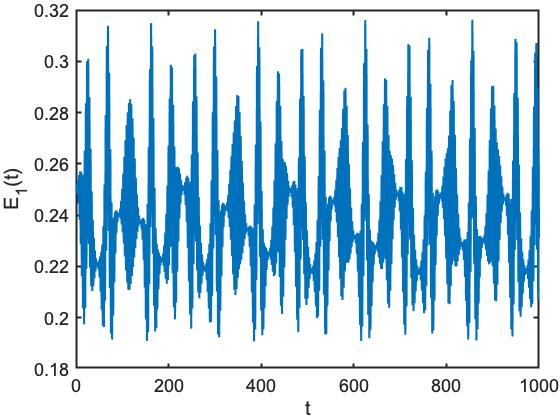}  \,  \includegraphics{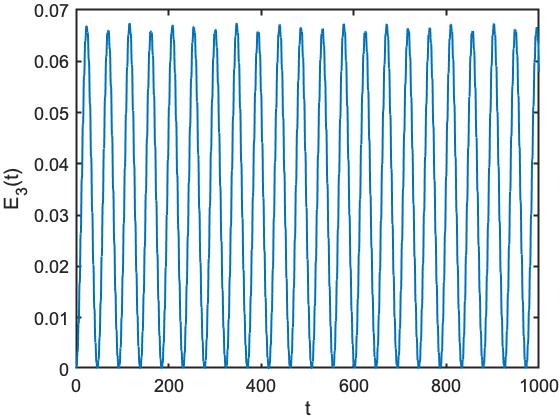}}
  \end{center}
 \caption{Interaction between optical and acoustical group in invariant manifold  $M$ corresponding with 4 particles.
The modes $x_1, x_2 $ are near $1:1$ resonance. We have in system \eqref{4part3dof} $ a= 0.01, x_1(0)=x_2(0)=0.5, x_3(0)=0$ 
and initial velocities zero. The instability of the solution in the optical group is indicated by the action 
$E_1(t)= 0.5( \dot{x}_1^2 + 2(1+a)x_1^2)$ (middle). 
Although far from resonance, the  low-frequency mode $x_3 $ is excited; figs left $x_3(t)$ and right 
$E_3(t)= 0.5 (\dot{x}_3^2 + 2ax_3^2)$.} 
\label{FPUfig2}
\end{figure} 
The eigenvalues and frequencies obtained from eq.~\eqref{eigalt} suggest that for mass $m$ large  we have
two groups 
of oscillators, one with frequency size close to $\surd 2$ and one with size $O(\surd a)$. There are indications in 
Bruggeman  and Verhulst~\cite{BV18} that in the case 
of a chain with 4 particles there exists significant interactions between the 2 groups. It turns out that in $\alpha$-chains the acoustical 
group can be strongly excited by the optical group.\\ 
We will clarify this interaction phenomenon in the case of 4n particles using the 4 particles invariant manifold $M$ that 
consists of the modes with frequencies $ \sqrt{1+a}, \sqrt{2}, \sqrt{2a}$. This submanifold corresponds 
with the 4 particles system described above.\\
As $0<a \ll 1$ there is actually no need for a scaling by small parameter  $\varepsilon $
 in this case. The corresponding equations of motion are (see Bruggeman and Verhulst \cite{BValt}): 
\begin{eqnarray} \label{4part3dof}
\begin{cases}
\ddot{x}_1 + 2(1+a) x_1 & =  2 \sqrt{a(1+a)} x_2 x_3,\\ 
\ddot{x}_2 + 2 x_2 & =  2 \sqrt{a(1+a)} x_1 x_3,\\
\ddot{x}_3 + 2a x_3 & =  2 \sqrt{a(1+a)} x_1 x_2. 
\end{cases} 
\end{eqnarray} 
The modes $x_1$ and $x_2$ are in a 
detuned $1:1$ resonance when choosing $0< a \ll 1$ . Consider the general position periodic solution  of the $1:1$ 
resonance of the $x_1, x_2$ modes, described in Bruggeman  and Verhulst~\cite{BV18}. A normal form approximation 
is $x_1(t)= r_0 \cos (\sqrt{2}t + \psi_0), x_1(t)= \pm x_2(t)$; the approximation is based on the equations for these modes 
to order  $O(a)$: 
\[ \ddot{x}_1 + 2x_1  =  2 \sqrt{a} x_2 x_3 + a \ldots,\,  \ddot{x}_2 + 2 x_2  =  2 \sqrt{a} x_1 x_3 + a \dots \] 
with $x_3$ varying on a long timescale. The asymptotic approximation with $x_1=x_2$ leads to a forced, linear 
equation for $x_3(t)$: 
\begin{equation} \label{slowx3} 
\ddot{x}_3 + 2a x_3  =  2 \sqrt{a} r_0^2 \cos^2 (\sqrt{2}t + \psi_0), 
\end{equation} 
with particular solution: 
\begin{equation} \label{acous}
x_3(t)= \frac{r_0^2}{2 \sqrt{a}}  - \frac{r_0^2}{8r_0^2 - 2 \sqrt{a}}\cos (2 \sqrt{2}t +2 \psi_0). 
\end{equation} 
To this  expression we have to add the homogeneous solution consisting of $ \cos (\sqrt{2a}t) $ and 
$\sin (\sqrt{2a}t )$. It is remarkable that the particular solution has a large amplitude, $O(1/ \sqrt{a})$, and period $\pi / \sqrt{2}$. The 
homogeneous solution has long period $\pi \sqrt{2}/\sqrt{a}$. 
We find 
that the ``acoustical mode'' $x_3$ is strongly excited; $x_1$ and $x_3$ are shown in fig.~\ref{FPUfig2} in the case of large mass 100. 

\section{Resonances induced by other mass ratio's} 
The classical FPU-chain and the chain with alternating masses are natural models of physical chains. It is clear from dynamical  
systems theory that resonances and symmetries play a fundamental part in all these model chains; see for instance 
Poincar\'e \cite{PMC} or Sanders et al.~\cite{SVM}. \\ 
Take for instance the classical FPU-chain with N=6; the 6 harmonic frequencies  
of system \eqref{FPUlin} are $1, \surd 3, 2, \surd 3, 1, 0$. As we know, both for $\alpha$- and $\beta$-chains $\bar{H}_3=0$ so the 
$1:2:1$ first order resonance is not effective because of symmetry; it might appear as a $2:4:2$ resonance at higher order. 
The $\surd 3 : \surd 3 = 1:1$ resonance plays a part for $\beta$-chains. 

A different choice of masses that would make the frequency spectrum of system 
\eqref{FPUlin} non-resonant would always have near-resonances as the rationals are dense in the set of real 
numbers. This would produce 
detuned resonances with behaviour related to exact resonance, so even in this case the analysis of Nishida \cite{N71} 
would not apply although his idea turns out to be correct. 
Thus it makes sense to explore systematically the kind of resonances that may arise in FPU-chains. As we shall see 
this leads to various applications. \\ 
The exploration of possible resonances was done by Bruggeman and Verhulst \cite{BV17} for the case of 4 particles 
leading to chains described 
by 3 dof. In Sanders et al.~\cite{SVM} ch.~10 a list 
of prominent Hamiltonian resonances in 3 dof is given for general Hamiltonians.  In {\em general} for 3 dof 
we have 4 first order resonances (active at $H_3$) and 12 second order 
resonances (active at $H_3 + H_4$). Considering system \eqref{FPUlin} for the special case of the FPU chains with arbitrary 
positive masses, we find that the first order 
resonance $1:2:2$ does not arise, of the 12 seond order resonances $1:1:1$ and $1:3:3$ are missing. The importance of the resonances 
that do arise is partly determined by the size of sets in the parameter space of masses. We present the results from Bruggeman and Verhulst \cite{BV17} where 
the sets in 3d-parameter space, the mass ratios of $(m_1, m_2, m_3, m_4)$, with active resonance are indicated between brackets: \\

{\bf First order resonance}\\
$1:2:1$ (4 points);\,$1:2:3$ (4 open curves);  $1:2:4$ (12 open curves). \\ 

{\bf Second order resonances} 
$$ \begin{array} {ll}
$1:1:3$ \,(4 \,points) & $1:2:5$\,(12\, open \,curves) \\
$1:2:6$\,(12\, open\, curves) & $1:3:4$\,(4 \,open\, curves) \\
$1:3:5$ \,(4 \,open \,curves) & $1:3:6$\, (12\,open\, curves) \\ 
$1:3:7$ \,(12 \,open \,curves) & $1:3:9$\,(12 \,open \,curves \\
$2:3:4$\, (2\, compact\,curves); & $2:3:6$\, (2\,compact \,curves)  
\end{array} $$
To determine the possible resonances for FPU chains with more than 4 particles is a formidable linear algebra and algebraic problem 
that has not been solved in generality. A general result from Bruggeman and Verhulst \cite{BV17} is that for 
$N \geq 4$ no mass distribution will 
produce the $N$ dof $1:1: \ldots : 1$ resonance. 
We will discuss some results that are known for the $1:2:3$ and $1:2:4$ resonances with 4 particles. The second 
order resonances are largely unexplored for FPU-chains. 

\subsection{The $1:2:3$ resonance} 
This resonance is of special interest as in this case for the general Hamiltonian chaos does not become exponentially 
small near stable equilibrium as 
$\varepsilon \rightarrow 0$ (see Hoveijn and Verhulst \cite{HV90}). In general the normal form of the $1:2:3$ resonance is not 
integrable; see Christov \cite{OC}. 
 However, symmetries may change the dynamics as is shown in systems with 4 
particles, see Bruggeman and Verhulst \cite{BV17} and below. \\

\begin{figure}[htp]
\begin{center}
\resizebox{12cm}{!}{
\includegraphics{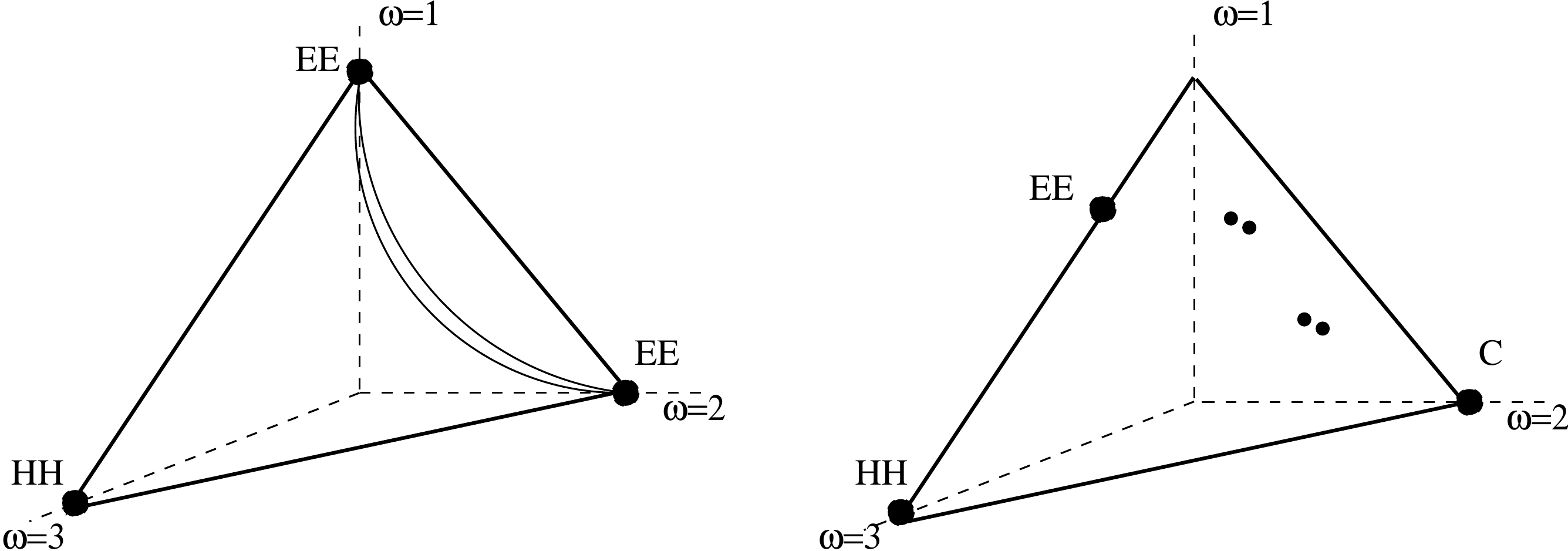}}
\end{center}
\caption{The $1:2:3$ resonance with action simplex of the symmetric case  $m_1 = m_3$ (left) 
and right a typical case with all masses different. Along the axes the actions 
 form a triangle for fixed values of $H_2$
which is an integral of the normal forms. The frequencies 
 $1, 2, 3$ indicate the 3  normal mode
positions at the vertices. The black dots indicate periodic solutions, the
indicated stability types are HH (hyperbolic-hyperbolic), EE
(elliptic-elliptic) and C (complex with real and imaginary parts nonzero). The two (roughly sketched) curves
connecting the 2 normal modes in the left simplex correspond for fixed energy 
with two tori consisting of periodic solutions, respectively with
combination angle $\chi=0$ and $\pi$. The tori break up into $4$ general
position periodic solutions if all masses are different. 
\label{FPUfig123}}
\end{figure} 
{\bf The symmetric  case of 4 particles $\alpha$-chains, $m_1=m_3$}\\
 Using integral \eqref{mom} and symplectic transformation we find the Hamiltonian:
 \begin{equation} \label{H123alpha}
 H(p, q)= \frac{1}{2} \sum_{j=1}^3 (p_j^2+ \omega_j^2 q_j^2) + \varepsilon(d_3q_1^2+ d_{10}q_2^2 +d_6q_3^2) q_3,\, 
 (\omega_1, \omega_2, \omega_3)= (3, 2, 1), 
 \end{equation} 
 with coefficients $d_3, d_6, d_{10} \neq 0$. The $(p_1, q_1)$ and the $(p_2, q_2)$ normal modes are exact periodic solutions in the 
 2 coordinate planes. Averaging-normalisation produces  in addition the $(p_3, q_3)$ normal mode periodic solution. 
 We find 3 
 integrals of motion of the normalised system so the normal form dynamics is integrable. The normal form system 
 contains only 
 one combination angle $\chi = \psi_1 - \psi_2- \psi_3$ producing for fixed energy families of periodic solutions (tori) in 
 general position. This is a degeneration in the sense described by Poincar\'e \cite{PMC} vol.~1, ch.~4. \\
 The stability of 
 the normal modes is indicated in fig.~\ref{FPUfig123}, left; the normal  2nd and 3rd modes ($\omega =2, 1$) are stable 
 with purely imaginary eigenvalues, the eigenvalues are coincident for the 2nd normal mode (Krein collision of eigenvalues). 
 The first mode ($\omega =3$) is unstable with real eigenvalues; In Bruggeman and Verhulst \cite{BV17} a 
 detailed description is  given of the motion of the orbits starting near the unstable normal mode ($\omega =3$). \\
 We will see that the case $m_1=m_3$ is structurally unstable, the dynamics changes drastically if all masses are different.\\ 
 
{ \bf The case of 4 particles $\alpha$-chains, all masses different}\\
 This case presents striking differences from the case with 2 masses equal, the symmetry is broken. We summarise:
 \begin{enumerate} 
 \item The 3rd normal mode ($\omega =1$) vanishes, the periodic solution shifts to the 2 dof subspace formed by the 
 first and 3rd mode; stability EE. 
 \item The second normal mode becomes complex unstable (C) by a Hamiltonian-Hopf bifurcation. In this case two 
 pairs of coincident imaginary eigenvalues (the case $m_1=m_3$) move into the complex plane. 
 \item The presence of a complex unstable periodic solution fits in the Shilnikov-Devaney scenario leading to chaotic 
 dynamics in the normal form, see Devaney \cite{D76} and  Hoveijn and Verhulst \cite{HV90}; 
 the normalised Hamiltonian is not integrable in this 
 case. Establishing chaos involves the presence of a horseshoe map. As this map is structurally stable, finding chaos 
 in the normal form, this chaos will persist in the original system. 
 \item The tori consisting of periodic solutions in the case $m_1=m_3$ break up into 4 periodic solutions at fixed energy. 
 
\end{enumerate} 

\subsection{The $1:2:4$ resonance} 
Work in progress for the FPU-chain with 4 masses in $1:2:4$ resonance can be found in Han{\ss}mann  et 
al.~\cite{HRV}; this analysis 
includes detuning. We mention some 
of the results in the case of opposing masses equal, $m_1=m_3$. 
\begin{enumerate} 
\item The case of 2 opposing masses equal induces a ${{\mathbb{Z}}_2}$ symmetry with as consequence that for both 
$\alpha$- and $\beta$-chains we have $\bar{H}_3 =0$. 
\item The normal form $H_2+ \bar{H}_4$ for the $\alpha$- and $\beta$-chains has 3 normal mode periodic solutions 
and is integrable. 
\item Normalisation to $H_6$ breaks the symmetry, only 2 integrals of the normalised Hamiltonian could be found. 
 \end{enumerate} 
 Interestingly the case of 2 adjacent equal masses produces different results; in this case $\bar{H}_3 \neq 0$, the 
 symmetry mentioned above is broken. \\
 The case of all masses different will be studied in a forthcoming paper.

\subsection{An application to cell-chains} 
\begin{figure}[ht]
 \begin{center}
  \resizebox{6cm}{!}{
   \includegraphics{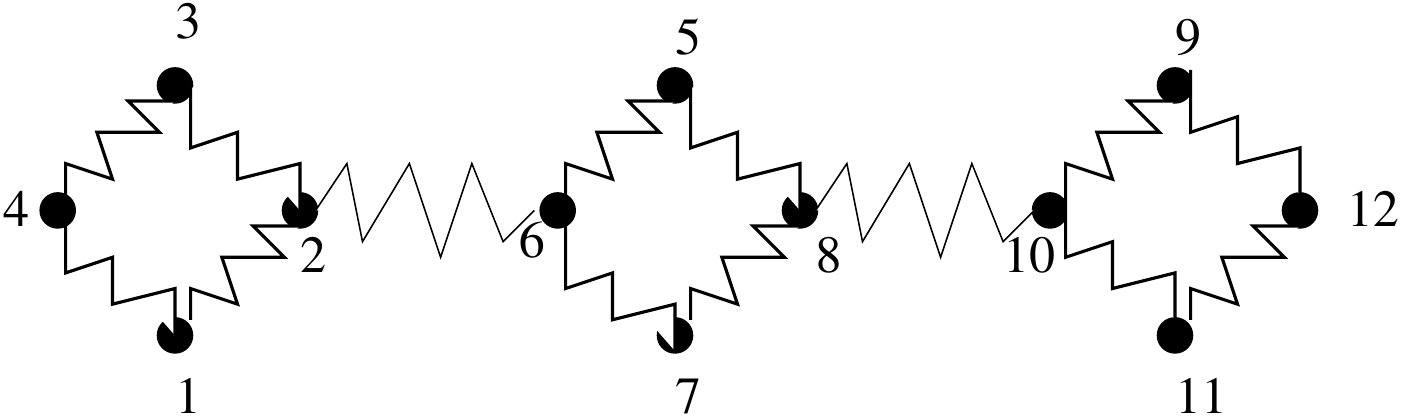}} 
  \end{center}
 \caption{A FPU cell-chain with 3 cells.} 
\label{FPUpict2}
\end{figure}

One can use low-dimensional FPU-chains as cells to form a new type of chain, see fig.~\ref{FPUpict2}. This is quite 
natural when thinking of interactions of molecules (a small group of connected oscillators) instead of atoms leading to a 
chain of  connected  near-neighbour interacting oscillators. A few examples of such cell-chains are discussed in 
Verhulst \cite{FVIJBC}. 

Consider cells consisting of a FPU-chain with 4 particles. As we have seen before the dynamics within each cell will 
strongly depend on the choice of the 4 masses. A second important aspect is how the cells are linked. Connecting cells 
by particles where stable periodic solutions dominate is expected to produce less transfer of energy than connecting by 
particles with more unstable periodic solutions and more active dynamics. 
Also the linking of cells will detune the resonances; this effect can be stronger if the FPU-chain is structurally unstable. 
We will show a few examples of transfer of energy for the simplest case of two connected cells. As the systems are 
Hamiltonian the phase-flow will always be recurrent but if the recurrence takes a long time this will indicate active 
but small transfer of energy between the cells with delayed recurrence. \\
Hamiltonian \eqref{H2cells} describes the interaction of 2 cells if $c_1 \neq 0$. 
 
\begin{equation} \label{H2cells} 
H(p, q)=  \sum_{j=1}^4 \left(\frac{m_j}{2} p_j^2+\frac{1}{2} (q_{j+1}-q_))^2) \right) + 
\sum_{j=5}^8 \left(\frac{m_{j-4}}{2 }p_j^2+\frac{1}{2} (q_{j+1}-q_j)^2 \right) + \frac{\varepsilon}{2}c_1(q_2-q_6)^2 +H_3,
\end{equation} 
with
\[ H_3 = \sum_{j=1}^8  \frac{\varepsilon}{3} (q_{j+1}-q_j)^3. \]
In the experiments we start with zero initial values 
in the 2nd cell, $q_j(0)=v_j(0)=0, j= 5, \ldots, 8$. If $c_1=0$ we have non-trivial dynamics and corresponding  
distance $d(t)$ to the initial values only in the first cell. Explicitly: 
\begin{equation} \label{recur2cells} 
d(t)= \sqrt{ \sum_{j=1}^8 [(q_j(t)-q_j(0))^2 + (v_j(t)-v_j(0))^2]}. 
\end{equation} 
The distance $d(t)$ can be used to consider recurrence to a $\delta$-neighbourhood of the initial values. 
An upper bound $L$ for the recurrence time has been given in Verhulst \cite{FVIJBC}. Suppose we consider a bounded 
Hamiltonian energy manifold with $N$ dof, energy value $E_0$ and Euclidean distance $d(t)$ of an orbit 
 to the initial conditions, than we have for the recurrence time $T_r$ to return in a $\delta$-neighbourhood of the 
 initial conditions an upper bound $L$ with: 
 \begin{equation} \label{boundL} 
 L= O \left( \frac{E_0^{N- 1/2}}{\delta^{2N-1}} \right). 
 \end{equation}  
 For one FPU-cell we have with reduction to 3 dof $L_1= E_0^{5/2}/ \delta^5$ and for 2 linked FPU-cells
$L_2= E_0^{13/2}/  \delta^{13}$. Of course, starting near a stable periodic solution or if there exist extra first integrals 
will reduce the recurrence time enormously. \\

\begin{figure}[ht]
 \begin{center}
  \resizebox{14cm}{!}{
   \includegraphics{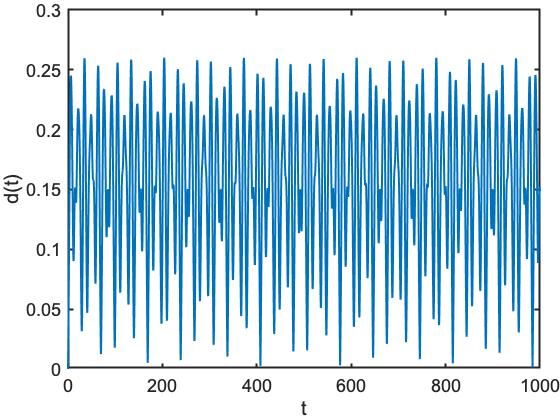} \,  \includegraphics{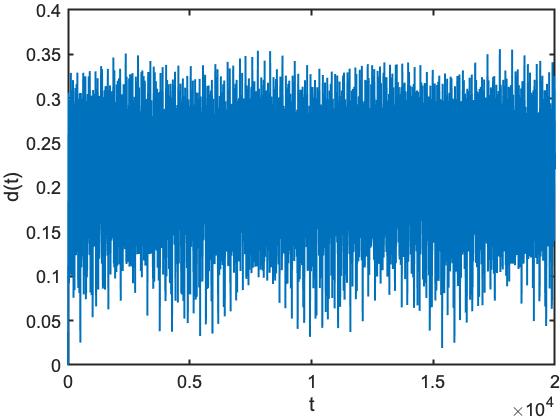}}
  \end{center} 
  \caption{The classical case $m=0.1$ with strong recurrence for 1 cell (roughly 100 timesteps if $\delta = 0.05 $) and 
  delayed recurrence for 2 cells (roughly 5000 timesteps). }
\label{FPUfig3}
\end{figure}

\begin{figure}[ht]
 \begin{center}
  \resizebox{14cm}{!}{
   \includegraphics{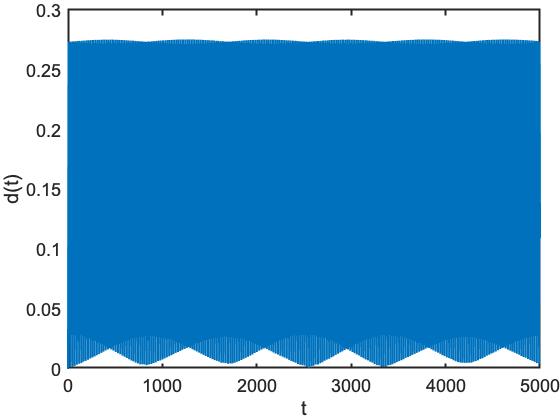} \,  \includegraphics{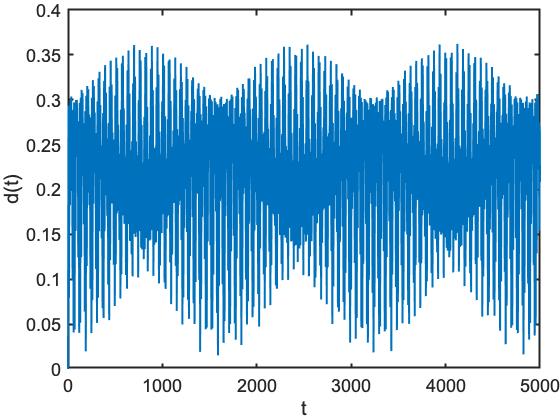}}
  \end{center}
\caption{The symmetric $1:2:3$ case $m_1=m_3$; the normal form is integrable, we have 
strong recurrence. Left one cell, with $\delta = 0.01$ roughly 800 timestep;, right 2 cells with $\delta = 0.05$  
roughly 1600 timesteps.  } 
\label{FPUfig4}
\end{figure}

\begin{figure}[ht]
 \begin{center}
  \resizebox{14cm}{!}{
   \includegraphics{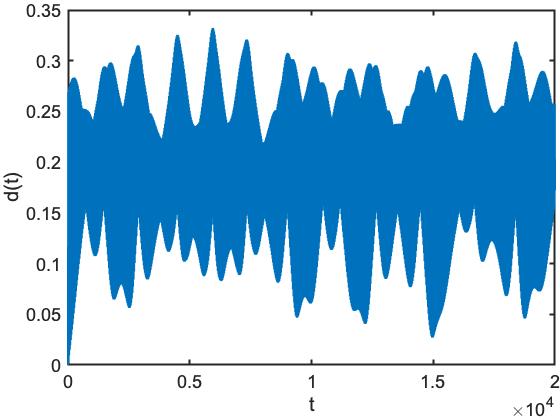} \,  \includegraphics{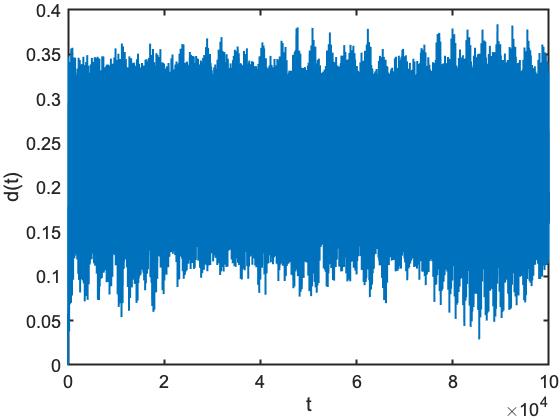}}
  \end{center}
 \caption{The chaotic $1:2:3$ cell-chain with already delayed recurrence in one cell; with $\delta = 0.05$ left 15000 
 timesteps; right for 2 cells we have to 
 integrate  nearly 90\,000 timesteps. } 
\label{FPUfig5}
\end{figure}

{\bf Numerical experiments}\\
We present numerical results for 3 cases with cells consisting of 4 masses: the classical FPU-chain with equal masses in fig.~\ref{FPUfig3} 
($m=0.1$ to have comparable timescales), 
the $1:2:3$ resonance case with symmetry induced by the choice $m_1=m_3$ in fig.~\ref{FPUfig4} and the less-balanced case 
of the $1:2:3$ resonance where the dynamics is chaotic, fig.~\ref{FPUfig5}. In each of the 3 cell-chains we have initial values 
$q_1(0)=0.05, q_2(0)=0.2, q_3(0)=0.05, q_4(0)=0.05, q_5(0)=q_6(0)=q_7(0)=q_8(0)=0.0$, initial velocities are
all zero. So we start in the first cell near the second normal mode plane. \\
As expected the recurrence times increase when adding one cell but most dramatically in the chaotic case. The inverse 
masses for fig.~\ref{FPUfig4} are $a_1= 0.0357143, a_2= 0.126804, a_3= 0.0357143, a_4= 0.301767$ 
(symmetric $1:2:3$ resonance with $m_1=m_3$) 
and for fig.~\ref{FPUfig5} $a_1= 0.00510292, a_2= 0.117265, a_3= 0.0854008, a_4= 0.292231$ 
(chaotic $1:2:3$ resonance). \\ 
In all these recurrence experiments with for instance $\delta = 0.1 $ or $\delta = 0.05 $ the recurrence times  
are definitely lower than the corresponding upper bound L given by eq.~\eqref{boundL}. \\
The numerics used {\sc Matlab} ode 78 with abs and rel error  $e^{-15}$.\\  

{\bf Acknowledgement}\\
Comments on earlier versions of this paper by Tassos Bountis and Roelof Bruggeman are gratefully acknowledged.


\begin{thebibliography}{99} 

\bibitem{BS12}
T. Bountis and H. Skokos, {\em Complex Hamiltonian Dynamics}, Springer (2012). 
 
\bibitem{BT} H.W. Broer and F. Takens, ({em Dynamical systems and chaos}, Applied Math. Sciences 172, Springer (2011).

\bibitem{BV18} R.W. Bruggeman and F. Verhulst, {\em Dynamics of a chain with four particles and nearest- neighbor 
interaction}, in Recent Trends in Applied Nonlinear Mechanics and Physics (M. Belhaq, ed.), CSNDD 2016, pp. 103-120, 
doi 10.1007/978-3-319-63937-6-6, Springer (2018).

\bibitem{BV17} Roelof Bruggeman and Ferdinand Verhulst, {\em The inhomogenous Fermi-Pasta-Ulam 
chain}, Acta Appl. Math. 152, pp. 111-145 (2017).

\bibitem{BValt} Roelof Bruggeman and Ferdinand Verhulst, {\em Near-integrability and recurrence in FPU 
chains with alternating masses}, J. Nonlinear Science 29, pp. 183-206, DOI 10.1007/s00332-018-9482-x (2019).

\bibitem{CRZ05} D.K. Campbell, P. Rosenau and G.M. Zaslavsky (eds.), {\em The Fermi-Pasta-Ulam Problem. The first 50 years.}
Chaos, Focus issue 15 (2005).

\bibitem{CS98} G.M. Chechin and V.P. Sakhnenko, {\em Interaction between normal modes in nonlinear dynamical systems 
with discrete symmetry. Exact results}, Physica D 117, pp. 43-76 (1998).

\bibitem{CNA} G.M. Chechin, N.V. Novikova and A.A. Abramenko, {\em Bushes of vibrational normal modes for 
Fermi-Pasta-Ulam chains}, Physica D 166, pp. 208-238 (2002).

\bibitem{CRZ} G.M. Chechin, D.S. Ryabov and K.G. Zhukov, {\em Stability of low-dimensional bushes of vibrational 
modes in the Fermi-Pasta-Ulam chains}, Physica D 203, pp. 121-166 (2005).

\bibitem{CEB} Christodoulidi, H., Efthymiopoulos, Ch. and Bountis, T. [2010]
{\em Energy localization on $q$-tori, long-term stability, and the
interpretation of Fermi-Pasta-Ulam recurrences}, Physical Review E 81,
6210.

\bibitem{OC} Ognyan Christov, {\em Non-integrability of first order
resonances in Hamiltonian systems in three degrees of freedom}, Celest.
Mech. Dyn. Astr. 112, pp. 149-167 (2012).

\bibitem{D76} R.L. Devaney, {\em Homoclinic orbits in Hamiltonian systems},
 J. Diff. Eqs. 21, pp. 431-438 (1976).

 \bibitem{E05} K. Efstathiou, {\em Metamorphoses of Hamiltonian systems with symmetries}, 
 Lecture Notes Math. 1864, Springer (2005). 

\bibitem{FPU55}E. Fermi, J. Pasta and S. Ulam, {\em Los Alamos Report
LA-1940}, in ``E. Fermi, Collected Papers'' 2, pp. 977-988 (1955).

\bibitem{Ford} J. Ford, {\em The Fermi-Pasta-Ulam problem: paradox turns discovery}, 
Physics Reports 213, pp. 271-310 (1992). 

\bibitem{G08} G. Galavotti (ed.) {\em The Fermi-Pasta-Ulam Problem: a status report}, Lecture Notes in 
Physics, Springer (2008). 

\bibitem{GGMV} L. Galgani, A. Giorgilli, A. Martinoli and S. Vanzini, On the problem of energy partition for large 
systems of the Fermi-Pasta-Ulam type: analytical and numerical estimates, Physica D 59, pp. 334-348 (1992).  

 \bibitem{H07} H. Han{\ss}mann, {\em Local and semi-local bifurcations in Hamiltonian dynamical systems}, 
 Lecture Notes Math. 1893, Springer (2007). 
 
 \bibitem{HRV} H. Han{\ss}mann, Reza Mazrooei-Sebdani and Ferdinand Verhulst, {\em The $1:2:4$ resonance 
 in a particle chain}, arXiv nr.~2002.01263, submitted to Indagationes Mathematicae (2020). 
 
\bibitem{H88} P.J. Holmes, J.E. Marsden and J. Scheurle,   {\em Exponentially small splittings of separatrices with application to KAM theory and degenerate bifurcations}, Contemp. Math. 81, pp. 213--244 (1988).

 \bibitem{HV90} I. Hoveijn, I. and F. Verhulst, {\em Chaos in the $1:2:3$- Hamiltonian normal form}, Physica D, 44, pp. 397-406 {1990}.

\bibitem{J91} E.A. Jackson,  {\em Perspectives of Nonlinear Dynamics}, 2 vols., 
Cambridge University Press, Cambridge  (1991). 

\bibitem{N71} T. Nishida, {\em A note on an existence of conditionally periodic oscillation in a one-dimensional  
anharmonic lattice}, Mem. Fac. Eng. Univ. Kyoto 33, pp. 27-34 (1971). 

\bibitem{PMC} Henri Poincar\'e, {\em Les M\'ethodes Nouvelles de la M\'ecanique
C\'eleste}, 3 vols., Gauthier-Villars, Paris (1892, 1893, 1899).

\bibitem{RV00} B. Rink and F. Verhulst, {\em Near-integrability of periodic FPU-chains}, Physica A 285, pp. 467-482 (2000).

\bibitem{BR}
B. Rink, {\em Symmetry and resonance in periodic FPU-chains}, 
Comm. Math. Phys. 218, pp. 665-685 (2001).

\bibitem{RC85}  D.L. Rod. and R.C. Churchill, {\em A guide to the H\'enon-Heiles Hamiltonian},  
Progress in Singularities and Dynamical Systems (S.N. Pnevmatikos, ed.), pp.  385-395, Elsevier (1985).

\bibitem{SVM} J.A. Sanders, F. Verhulst and J. Murdock, {\em Averaging methods in nonlinear dynamical 
systems} 2nd ed., Appl. Math. Sciences 59, Springer, New York etc., (2007).

\bibitem{FV85} Ferdinand Verhulst, {\em Nonlinear differential equations and dynamical systems} 2nd ed.,
Springer, New York etc., (2000). 
 
 \bibitem{FVIJBC} Ferdinand Verhulst, {\em Near-integrability and recurrence in FPU cells},
 Int. J. Bif. Chaos 26, nr 14, DOI: 10.1142/S0218127416502308 (2016). 
 
 \bibitem{FV18} Ferdinand Verhulst, {\em Linear versus nonlinear stability in Hamiltonian systems}, Recent trends in Applied Nonlinear Mechanics and Physics, Proc. in Physics 199 (M. Belhaq, ed.) pp. 121-128, (2018) Springer, DOI 10.1007/978-3-319-63937-6-6.

\bibitem{Z}
G.M. Zaslavsky, {\em The physics of chaos in Hamiltonian systems}, Imperial College Press 
(2nd extended ed.) (2007). 


\end{thebibliography}
\end{document}